\documentclass{elsart}
\usepackage{amssymb}
\usepackage{amsmath}
\usepackage{graphicx}
\newcommand{\bmath}[1]{\mbox{\boldmath ${#1}$}}
\newcommand{\dd}{\mbox{\rm d}}
\begin{document}

\begin{frontmatter}

\title{\large{\bf Spin-triplet final-state dominance in the
$\bmath{pp\to pn\,\pi^+}$ reaction at 492~MeV}}

\author[gatchina]{V.~Abaev},
\author[juelich]{M.~B\"uscher\thanksref{CA}},
\author[juelich,dubna]{S.~Dymov},
\author[juelich]{M.~Hartmann},
\author[dubna,tiflis]{A.~Kacharava},
\author[dubna]{V.I.~Komarov},
\author[gatchina]{V.~Koptev},
\author[dubna]{V.~Kurba\-tov},
\author[gatchina]{S.~Mi\-kir\-tychiants},
\author[juelich,gatchina]{M.~Ne\-ki\-pe\-lov},
\author[juelich,dubna]{A.~Petrus},
\author[juelich]{H.~Str\"oher},
\author[dubna]{Yu.N.~Uzikov\thanksref{WTZ}},
\author[london]{C.~Wilkin},
\author[dubna]{S.~Yaschenko}

\address[gatchina]{High Energy Physics Department, Petersburg Nuclear Physics
         Institute, 188350 Gatchina, Russia}
\address[juelich]{Institut f\"ur Kernphysik, Forschungszentrum J\"ulich,
         D-52425 J\"ulich}
\address[dubna]{Laboratory of Nuclear Problems, Joint Institute for Nuclear
         Research, Dubna, 141980 Dubna, Moscow Region, Russia}
\address[tiflis]{High Energy Physics Institute, Tbilisi State University,
         University St.\ 9, 380086 Tbilisi, Georgia}
\address[london]{Physics Department, UCL, Gower Street,
         London WC1 6BT, England} 
\thanks[CA]{Corresponding author. E-mail address: m.buescher@fz-juelich.de}
\thanks[WTZ]{Supported by BMBF, Germany (WTZ grant KAZ 99/001)}

\begin{abstract}
  The near-forward cross section for the $pp\to pn\pi^+$ reaction 
  has been measured at 492~MeV by using the large acceptance ANKE 
  magnetic spectrometer placed at an internal target position of the 
  storage ring COSY-J\"ulich. Protons and pions emitted near zero
  degrees were detected in coincidence, and those with
  $\theta_{\pi}\leq 2^{\circ}$ and $\theta_p\leq 2.5^{\circ}$ were
  subjected to detailed analysis. Under these conditions, the
  excitation energy in the $np$ system was below 3~MeV over the
  measured momentum range. This is the region of the $np$
  final-state-interaction peak, which was measured with a
  resolution of a fraction of an MeV. The shape of the peak allows one 
  to conclude that the fraction of final spin-singlet $np$ pairs is 
  below about 10\%. By using the results of scattering theory, this 
  limit is confirmed through a comparison with the cross section for 
  $pp\to d\pi^+$. The smallness of the singlet contribution is 
  consistent with trends seen in lower energy data.
\end{abstract}

\end{frontmatter} 
\newpage
\baselineskip 4ex

The two-body $pp\to d\pi^+$ reaction has long been used as a test
bed for both experimental measurements and theoretical models of 
intermediate energy pion production~\cite{SAID}. Unfortunately,
there is no $^1S_0$ bound state in the $S$-wave spin-0 $np$ system,
analogous to the $^3S_1$ deuteron bound state in the spin-1 system,
that would enable a similar study of the other isospin channel. 
On the other hand, there is a virtual (antibound) $^1S_0$ state 
very close to threshold at an $np$ excitation energy
$E_{np}\approx -0.07$~MeV. Both this and the deuteron pole at
$-2.22$~MeV induce strong $S$-wave final-state-interaction (fsi)
peaks in the $pp\to pn\pi^+$ reaction leading to spin-singlet
and triplet $np$ channels. To simulate a two-body reaction, the
excitation energy in the $np$ system should be confined to within
a few MeV. Furthermore, the singlet and triplet contributions should
be cleanly separated. This should in fact be possible with good
energy resolution because the fsi peak widths are proportional to
the (real or virtual) binding energies. The width of the singlet
peak would therefore be a fraction of an MeV as compared to the
2-3~MeV of the triplet.

Inclusive $pp\to \pi^+X$ reactions at intermediate energies show an
$np$ enhancement at the edge of phase space, but the energy resolution
achieved and the contamination from the much stronger $pp\to d\pi^+$
channel makes it very difficult to study the fsi region at low
$E_{np}$~\cite{Falk}. The data nevertheless suggest that the triplet
fsi provides the major signal~\cite{FW2}. To avoid feed-through from
the $d\pi^+$ final state, it is useful to measure the proton and pion
of the $pn\pi^+$ channel in coincidence and this can also lead to
better determination of $E_{np}$. Such an experiment was carried out
at LAMPF at 800~MeV~\cite{HG}. Although fsi peaks were observed, the
number of $E_{np}$ points in the very low energy region was limited,
as was the resolution, and so it was not possible to isolate a narrow
singlet fsi peak from a broader triplet one.  Nevertheless their
model-dependent analysis suggested that the singlet fsi represented a
very large fraction of the total strength at all angles measured. We
report here on the results of an exclusive $pp\to pn\pi^+$ measurement
at $T_p=492$~MeV, where both the proton and pion were detected close
to the forward direction. The resolution of a fraction of an MeV in
$E_{np}$ allows us to separate contributions from the singlet and
triplet fsi peaks and our results suggest that the singlet fraction is
below 10\%. This upper bound is confirmed by comparing the magnitude
of our cross section with that for $pp\to d\pi^+$ using general
scattering theory results~\cite{UW}.

The experiment was carried out at the ANKE spectrometer
placed inside the storage ring of the proton synchrotron
COSY-J\"ulich. The main ANKE components are shown in
Fig.~1, but see Ref.~\cite{ANKE} for a detailed description.
Foil targets made of CH$_2$ (polyethylene) were used in
the experiment and the background from carbon was measured using
a polycrystalline diamond target. The spectrometer dipole D2
separates positively charged ejectiles from the circulating COSY
beam and allows one to determine their momenta at the target over
a wide momentum range.  The horizontal angular acceptance of
$|\vartheta_{\mathrm H}|\leq 12^{\circ}$ is mainly fixed by the
allowed start-stop combinations of the on-line TOF-trigger system.
The vertical angular acceptance is in the range
$|\vartheta_{\mathrm V}|\leq 7^{\circ} (3.5^{\circ})$ for the lowest
(highest) detectable ejectile momenta.

\input epsf
\begin{figure}[ht]
  \begin{center}
    \mbox{\epsfxsize=4in \epsfbox{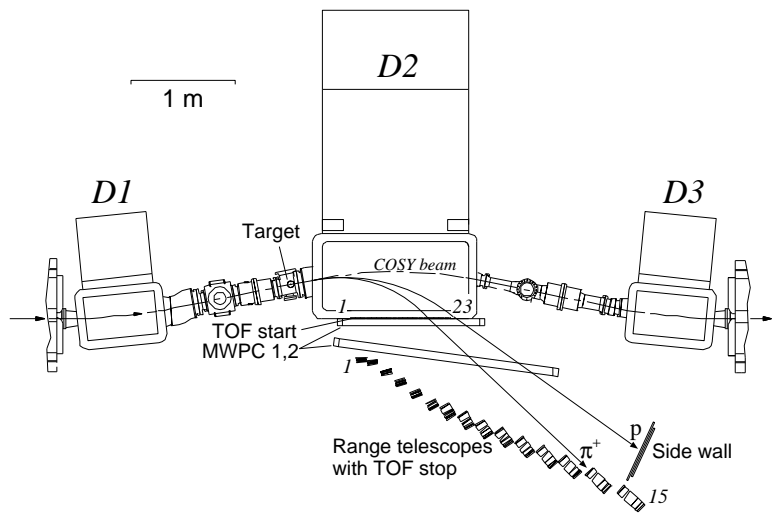}}
    \caption{Top view of the ANKE spectrometer and detectors used for
      the FSI studies.}
  \end{center}
  \label{fig:anke}
\end{figure}

The detection system~\cite{K_NIM} comprises two
multi-wire proportional chambers (MWPC 1, 2 in Fig.~1)
for track and momentum reconstruction, 23 scintillation counters
close to the vacuum window of D2 for time-of-flight start
measurements and 15 range telescopes located along the focal surface
of D2. The side wall counters allow larger ejectile momenta to be
measured. The identification of pions and protons is achieved
\textit{via} TOF and energy loss measurements. Background
originating from the pole shoes of D2 is suppressed with the help of
the track information from the MWPC's.  The field strength of
$B=0.861$~T in D2 was chosen so that the pions used in the analysis
were detected and identified in telescope \#14 and the protons in
the side-wall counters.

The spectrum of missing-mass-squared for the $p(p,p\pi^+)X$ reaction
on a CH$_2$ target, shown in Fig.~2 for the full ANKE angular
acceptance, displays a prominent neutron peak, with a standard
deviation of $\sigma(m_x^{\,2})= 0.0034\,m_n^{\,2}$, which is not
present for the C target. The integrated luminosities for the two
targets are different but, when the relative CH$_2$/C normalisation
is evaluated from the spectra at high missing masses, it is found that
the neutron peak sits on a carbon background of a few percent, which
can easily be subtracted. Fits to the position of the deuteron and
neutron peaks determined the proton beam energy to be
$T_p = (492\pm 1.7)$~MeV.

\input epsf
\begin{figure}[h]
\begin{center}
\mbox{\epsfxsize=3in \epsfbox{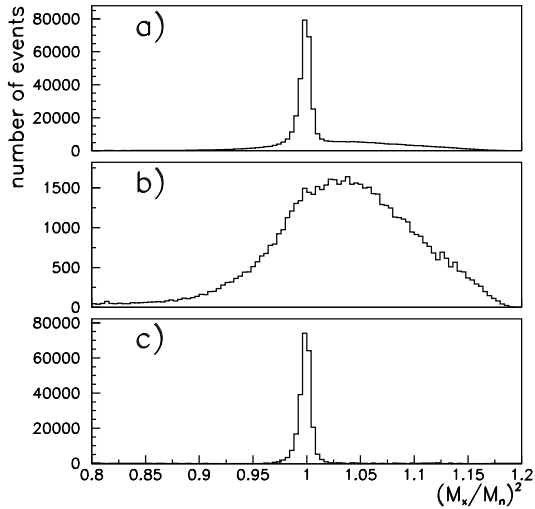}}
\caption{Spectrum of the square of the missing mass for the
$p(p,p\pi^+)X$ reaction, in units of the neutron mass, for: a) CH$_2$;
b) C; c)  CH$_2 - $C targets, normalised to the same luminosity. By
comparing the spectra at high masses, the relative luminosity
$L($C$)/L($CH$_2)$ is deduced to be $0.264\pm 0.020$.}
\end{center}
\label{fig2}
\end{figure}

In order to obtain clean samples of events with low $np$ excitation
energies, software cuts were imposed so as to extract two groups of
events. Both of these have pion angles $\theta_{\pi}\leq 2.0^{\circ}$,
but the first group includes protons with angles
$\theta_p\leq 2.0^{\circ}$, whereas the proton cut was extended to
$2.5^{\circ}$ for the second set. After subtracting the C background,
the number of events remaining in the two cases was 1312 and 2008
respectively. The ratio follows that of the proton solid angle, the cut
in the pion angle being of no importance here.

In the raw proton momentum spectra shown in Fig.~3, the carbon
background with our cuts is below 3\%. The central $np$ excitation
energies are indicated and, from this scale, it is seen that our
events generally correspond to $E_{np}\leq 3$~MeV, which means that
the fsi region is well covered. Because of the limited statistics, we
have summed all events over the whole angular ranges and this introduces
an uncertainty of $\Delta E_{np}\approx 0.55\sqrt{E_{np}}$~MeV. This is
far greater than the intrinsic resolution of the system which, through
kinematic fits to the $pp\to pn\pi^+$ reaction, is about
$\sigma=160$~keV. Thus the width of any spin-singlet contribution
in Fig.~3 is determined by the angular integration rather than the
natural width or the intrinsic energy resolution of the apparatus.

\input epsf
\begin{figure}[ht]
\begin{center}
\mbox{\epsfxsize=4in \epsfbox{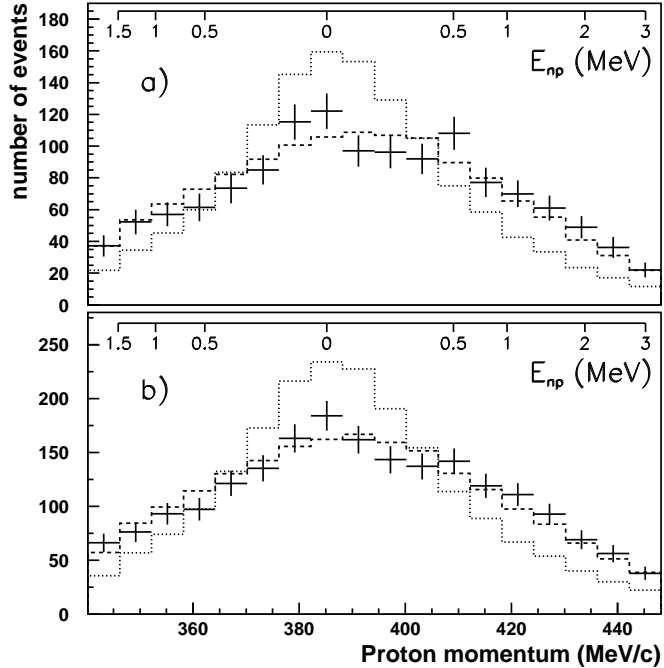}}
\caption{Raw proton-momentum spectra of $pp\to pn\pi^+$ events 
  (points with statistical errors), for events with
  $\theta_{\pi}\leq 2^{\circ}$. All the histograms are normalised to
  the total number of events. The relative $np$ energy is indicated
  for protons and pions emitted at $0^{\circ}$. Figure a) is for
  $\theta_p\leq 2^{\circ}$ and b) for $\theta_p\leq 2.5^{\circ}$.
  The dashed histogram is a Monte Carlo simulation assuming pure
  spin-triplet final state ($\xi=0$). Fits to the fsi formula of
  eq.~(\protect\ref{Vladimir}) are consistent with this, giving
  $\xi = -.008 \pm .038$ for case a) and $\xi = -.008 \pm .036$ for
  b. A statistical mixture of spin states, \textit{i.e.}\ $\xi=0.25$,
  leads to the dotted histograms with $\chi^2/$\textit{ndf} =151.2/17
  for a) and $\chi^2/$\textit{ndf} =178.2/17 for b).
  \vspace{5mm}\label{fig3}}
\end{center}
\end{figure}

With the angular cuts we have imposed, the proton measurement
efficiency is about 60\% at small proton momenta ($\leq$~350~MeV/c),
constant at about 99\% for momenta between 350~MeV/c and 410~MeV/c,
but falls rapidly thereafter. Taking this and other effects into
account in a detailed Monte Carlo study, we show in Fig.~3 a
simulation where we have assumed that the cross section is
proportional to three-body phase space times essentially a pure
spin-triplet fsi factor. The significance of the excellent
agreement with the shape of the spectrum will be crucial for the
later discussion. One of the advantages of our experimental set-up is
that we also have a measurement of the corresponding pion momentum
spectrum. However, within the current precisions, the singlet/triplet
ratio is better determined from the proton spectrum~\cite{AKS}.

The absolute normalisation of the $pp\to pn\pi^+$ cross section is
achieved by comparing the $pp\to d\pi^+$ events, measured in parallel,
to standard cross section compilations~\cite{SAID}. Because of the
dominance of the $d\pi^+$ final state, the reaction could be identified
with sufficient precision without having to detect the deuteron, thus
avoiding uncertainties arising from deuteron break-up in the counters
\textit{etc.} The contamination from the carbon background is here
larger ($\approx 23\%$) than for three-body events, but it is easy
to correct for this using the normalised results from the carbon
target discussed earlier. Our results are presented in Fig.~4, where
only statistical errors are shown. The principal systematic effects
are:
\begin{itemize}
\item
Uncertainty in the number of $pp\to d\pi^+$ events, due to
the tail of pions from the $pp\to pn\pi^+$ reaction where the proton
escapes detection: $\approx 7\%$.
\item
Uncertainty in the ratio of C/CH$_2$ integrated luminosities:
$\approx 2.5\%$.
\item
Uncertainty in the effective proton solid angle: $\approx 1.5\%$.
\item
Uncertainty in the SAID  solution for $pp\rightarrow d \pi^+$, used
for normalisation: $\approx 2.0\%$.
\end{itemize}
Summing these uncertainties quadratically, we arrive at an estimate of
the total systematic error of $\approx 8\%$.

\input epsf
\begin{figure}[h]
\begin{center}
\mbox{\epsfxsize=4in \epsfbox{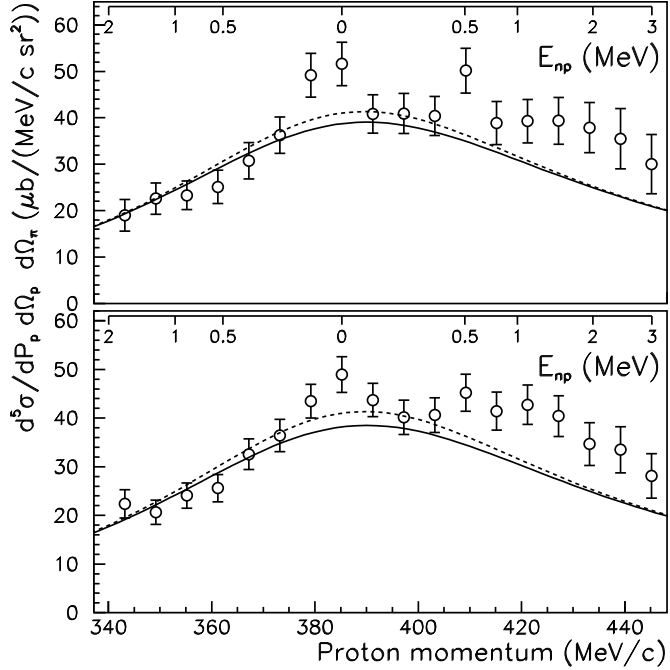}}
\caption{Five-fold differential cross section for the $pp\to pn\pi^+$
  reaction as a function of the measured proton momentum for events
  with $\theta_{\pi}\leq 2^{\circ}$. The $np$ excitation energy is
  indicated for protons and pions emitted at $0^{\circ}$. The upper
  figure is for $\theta_p\leq 2^{\circ}$ and the lower for
  $\theta_p\leq 2.5^{\circ}$.  The solid curves correspond to
  predictions for the spin-triplet contribution to the fsi peak,
  obtained from eq.~(\protect\ref{yuri}) using $pp\to d\pi^+$ input
  data from the SAID SP96 solution~\protect\cite{SAID}, after
  averaging over the angular acceptance. The dashed curves are raw
  predictions without averaging.  The estimates are expected to be
  only weakly model-dependent~\protect\cite{UW}.
  \vspace{5mm}\label{fig4}}
\end{center}
\end{figure}

Since the $pp\to pn\pi^+$ events of Fig.~3 and Fig.~4 are all
concentrated in the region $E_{np}\leq 3$~MeV, we can make an
estimate of the relative amounts of spin-singlet and -triplet final
states using the fact that any singlet fsi peak should be so much
narrower than the triplet. We assume a variant of the
Goldberger-Watson fsi formula~\cite{GW}
\begin{equation}
\label{GW}
\textrm{FSI}_i(k) = N_i\,\frac{(k^2+\beta_i^{\,2})}{(k^2+\alpha_i^{\,2})}\:,
\end{equation}
where the $\alpha_i$ and $\beta_i$ are determined from low energy
$np$ scattering data~\cite{Machleidt}. The triplet and singlet
parameters are $\alpha_t=0.232$~fm$^{-1}$, $\alpha_s=-0.040$~fm$^{-1}$,
$\beta_t=0.91$~fm$^{-1}$, and $\beta_s=0.79$~fm$^{-1}$. It should be
noted that the numerators in eq.~(\ref{GW}) vary very little over our
restricted $E_{np}$ range, so that the fits are weakly dependent upon
the values of the $\beta_i$. The overall constant $N_i$ is determined
by the condition that the integral of $\textrm{FSI}_i(k)$ over $k$ for
$E_{np}\leq 3$~MeV is unity, which requires that $N_t=0.104$~fm and
$N_s=0.0297$~fm. We then fit our data with the form
\begin{equation}
\label{Vladimir}
\frac{\dd^5\sigma_t(pp\to pn\pi^+)}{\dd p_p\,\dd\Omega_p\,\dd\Omega_\pi}
\propto (\textrm{Kinematic factor})
\times \left[\xi\,\textrm{FSI}_s(k) + (1-\xi)\,\textrm{FSI}_t(k)\right]\:,
\end{equation}
so that $\xi$ has the physical significance of being the fraction of 
the cross section leading to singlet $np$ states for $E_{np}\leq 3$~MeV.

When this form is passed through the Monte Carlo simulation and
normalised to the data in Fig.~3, an excellent fit is achieved with
$\xi = -.008 \pm .038$ and $\chi^2/\textit{ndf} = 18.2/17$ for group a)
and $\xi= -.008 \pm .036$ and $\chi^2/\textit{ndf} = 15.9/17$ for group
b). Effects of smearing the predictions over the spread in $E_{np}$
induced by the angular acceptance and momentum bite \textit{etc.}\ have
been taken into account. A value of $\xi=0.25$, which would correspond
to a purely statistical factor, gives excess of events in the region
around zero excitation energy and shortage at the right tail of the
spectrum. This increases the $\chi^2/ndf$ by about 10 units. Thus we
can assert purely from the shape of the fsi peak that the singlet
contribution to the cross section is likely to be below $\approx$~10\%.

A second proof that the spin-singlet fraction is small can be found
from the normalisation rather than the shape of the measured
differential cross section. The F\"aldt-Wilkin extrapolation theorem
links the $np$ scattering wave function to the deuteron bound-state
function, independently of the form of the $np$
potential~\cite{FW1,FW2}, and this allows one to predict the $S$-wave
spin-triplet contribution to the $pp\to pn\pi^+$ cross section at low
$E_{np}$ in terms of that for $pp\to d\pi^+$~\cite{UW}. To the extent
that deuteron $D$-state effects can be neglected, 
\begin{equation}
\label{yuri}
\frac{\dd^5\sigma_t(pp\to pn\pi^+)}{\dd p_p\,\dd\Omega_p\,\dd\Omega_\pi}
= (\textrm{Kinematic factor})\times \frac{1}{(k^2+\alpha_t^2)}\times
\frac{\dd\sigma}{\dd\Omega_{\pi}}(pp\to d\pi^+)\:.
\end{equation}
This relation does not require explicit knowledge of the pion production
operator, merely that it be of short range. It describes well~\cite{UW}
the fsi region of the LAMPF $pp\to pn\pi^+$ data taken at a variety of
angular configurations~\cite{HG}, though it should be stressed that the
number of experimental points in the fsi peaks was small. 

Because eq.~(\ref{yuri}) has the same dominant fsi factor as that of the
spin-triplet Goldberger-Watson factor of eq.~(\ref{Vladimir}), it is no
surprise that the shape of the differential cross section as a function
of the final proton momentum shown in Fig.~4 is largely reproduced. To
within possible corrections of the order of the deuteron $D$-state
probability ($\approx 6\%$), this approach provides a robust lower
bound on the three-body cross section. Although the curves pass through
the points to the left of the peak, they are slightly low compared
to the data on the right. This is compatible with theoretical 
and experimental uncertainties, and so the normalisation of the data 
also shows that any singlet contribution must be below about 10\%.

The inclusive TRIUMF $p(p,\pi^+)X$ data~\cite{Falk} indicate
significant contributions from $np$ final states other than triplet
$S$-wave, but their excitation energies are such that $P$-waves cannot
be neglected. The exclusive $p(p,p\pi^+)n$ measurement carried out at
CELSIUS at 400~MeV~\cite{Betsch} was analysed as for a single-arm
experiment by integrating over the final proton momentum as well as
over the pion angle. When this is done, there is an indication of 
contributions other than spin-triplet $S$-waves, though there is only
one high point in the $E_{np}\leq 3$~MeV range.

Given that the singlet fraction is so small at intermediate energies, 
this suggests that one should measure its production directly by 
studying the $pp\to pp\pi^0$ reaction, which could be done in parallel 
with $pn\pi^+$ detection at ANKE. The reaction has already been 
investigated at CELSIUS up to 425~MeV and values of the differential 
cross section quoted separately for $E_{pp} \leq 2.6$~MeV~\cite{Jozef}. 
Though there is no similar detailed measurement of the 
$pp\to pn\pi^+$ cross section with which to compare the CELSIUS
$\pi^0$ data, it is possible to estimate the spin-triplet contribution 
quite reliably from that of $pp\to d\pi^+$ by employing the 
F\"aldt-Wilkin theorem~\cite{FW2}. If then the cross section ratio is 
extrapolated to 492~MeV, it indicates that the singlet fraction should 
be below about 1\%. This very low value is due, in part, to $\pi^+$
production being maximal in the forward direction~\cite{SAID}, whereas
the $\pi^0$ are produced preferentially at $90^{\circ}$~\cite{Jozef}.
Thus the singlet fraction at 492~MeV would be expected to be much
larger for an experiment carried out at $90^{\circ}$.

In summary, we have measured the exclusive $pp\to pn\pi^+$ cross
section reaction near the forward direction with a resolution in
$E_{np}$ superior to that of previous works. This allows us to put
an upper bound on the proportion of singlet $np$ final states, which
is confirmed independently by comparing with $pp\to d\pi^+$ results. 
It is hoped to continue the investigation at other energies,
studying simultaneously the $pn\pi^+$ and $pp\pi^0$ channel.

The results presented here were obtained during the ANKE commissioning
experiments. We would like to thank the colleagues who helped to
realise the ANKE spectrometer, in particular O.W.B.~Schult,
K.~Sistemich and R.~Maier, and the many former Diploma and Ph.D.\ 
students. We express our thanks to the COSY team for its dedication
and very smooth implementation of ANKE into the accelerator ring.
During the measurements, support from the ``Zentrallabor f\"ur
Elektronik'' (ZEL) of the FZJ, in particular W.~Erven, was
indispensable.



\newpage
\begin{thebibliography}{99}
%
\bibitem{SAID} R.A. Arndt et al., Phys.\ Rev.\ C~48 (1993) 1926;\\
 \verb+http://gwdac.phys.gwu.edu/+
%
\bibitem{Falk} W.R.~Falk et al., Phys.\ Rev.\ C~32 (1985) 1972; 
R.G.~Pleydon et al., Phys.\ Rev.\ C~59 (1999) 3208.
%
\bibitem{FW2} A.~Boudard, G.~F\"aldt and C.~Wilkin, 
Phys.\ Lett.\ B~389 (1996) 440.
%
\bibitem{HG} J.~Hudomalj-Gabitzch et al., Phys.\ Rev.\ C~18 
  (1978) 2666.
%
\bibitem{UW} Yu.N.~Uzikov and C.~Wilkin, Phys.\ Lett.\ B 511 (2001) 191.
%
\bibitem{ANKE} S.~Barsov et al., Nucl.\ Instr.\ Meth.\ A~462 (2001) 364.
%
\bibitem{K_NIM} M.~B\"uscher et al., Nucl.\ Instr.\ Meth.\ A (in press).
%
\bibitem{AKS} V.~Abaev, V.~Koptev and H.~Str\"oher, {\it $\pi^+p$ 
correlation at $0^{\circ}$ and $np$ final state interaction in
$pp\to \pi^+ pn$ reaction at 495~MeV}, preprint PNPI-2001 No.~2403.
%
\bibitem{GW} M.L.~Goldberger and K.M.~Watson, {\it Collision Theory}
(John Wiley, N.Y.) 1964. 
%
\bibitem{FW1} G.~F\"aldt and C.~Wilkin, Physica Scripta 56 (1997) 566.
%
\bibitem{Betsch} A.~Betsch et al., Phys.\ Lett.\ B~446 (1999) 179.
%
\bibitem{Jozef} R.~Bilger et al., Nucl.\ Phys.\ A (in press).
%
\bibitem{Machleidt} R.~Machleidt, Phys.\ Rev.\ C~63 (2001) 024001.

\end{thebibliography}
\end{document}